\documentclass[twocolumn,preprintnumbers,amsmath,amssymb]{revtex4}

\usepackage{graphicx}
\usepackage{bm}
\usepackage{color}
\usepackage{amsmath}
\usepackage{amsfonts}
\usepackage{amssymb}
\usepackage{epstopdf}
\usepackage{ulem}

\begin{document}

\preprint{APS/123-QED}

\title{Field-orientation dependence of quantum phase transitions in the $\bm{S=1/2}$ triangular-lattice antiferromagnet Ba$_3$CoSb$_2$O$_9$}

\author{Kazuki Okada$^1$}
\author{Hidekazu Tanaka$^1$}
\email{tanaka.h.ag@m.titech.ac.jp}
\author{Nobuyuki Kurita$^1$}
\author{Daisuke Yamamoto$^2$}
\author{Akira Matsuo$^3$}
\author{Koichi Kindo$^2$}

\affiliation{
$^1$Department of Physics, Tokyo Institute of Technology, Meguro-ku, Tokyo 152-8551, Japan\\
$^2$Department of Physics, College of Humanities and Sciences, Nihon University, Setagaya-ku, Tokyo 156-8550, Japan\\
$^3$Institute for Solid State Physics, The University of Tokyo, Kashiwa, Chiba 277-8581, Japan
}
\date{\today}
 
\begin{abstract}
Ba$_3$CoSb$_2$O$_9$ approximates the two-dimensional spin-1/2 triangular-lattice Heisenberg antiferromagnet. This compound displays magnetic-field-induced quantum phase transitions, including the 1/3-magnetization-plateau, but its magnetization processes for the magnetic field $H$ parallel and perpendicular to the $c$ axis are different due to the weak easy-plane anisotropy and the weak interlayer antiferromagnetic exchange interaction. To elucidate how the quantum phase transitions change between these two field directions, we measured the field-angle dependence of the magnetization process in Ba$_3$CoSb$_2$O$_9$ using pulsed high magnetic fields. We compared obtained magnetic field--field angle phase diagram with those obtained by the large-size cluster mean-field method combined with a scaling scheme and the semiclassical theory. We also found a narrow 1/3-magnetization plateau and a high-field transition with a small magnetization jump for $H\,{\parallel}\,c$, not observed in the previous studies.
  
\end{abstract}


\maketitle

\section{Introduction}

Usual phase transitions exemplified by the solid-liquid and ferromagnetic-paramagnetic transitions are induced by temperature variations. The driving forces are thermal fluctuations. On the other hand, quantum phase transitions (QPTs) are transitions between different quantum ground states caused by the continuous variation of interactions, which can be tuned by the magnetic field, pressure, etc. QPTs are driven by the quantum fluctuation~\cite{Sachdev,Vojta}. The $S\,{=}\,1/2$ triangular-lattice Heisenberg antiferromagnet (TLHAF) exhibits typical QPTs. The ground state of the two-dimensional (2D) $S\,{=}\,1/2$ TLHAF with the nearest-neighbor exchange interaction at zero magnetic field has been considered to be an ordered state with the 120$^{\circ}$ structure~\cite{Huse,Jolicoeur,Bernu,Singh,White,Gotze,Ghioldi}. Although the ground state is qualitatively the same as the classical ground state, the sublattice moment is reduced to about 40\,\% of the full moment due to the quantum fluctuation~\cite{Huse,Jolicoeur,Bernu,Singh,White,Gotze,Ghioldi,Li2}.

In a magnetic field, an up-up-down (UUD) spin structure is stabilized in a finite field range by the quantum fluctuation. Consequently, the 2D $S\,{=}\,1/2$ TLHAF exhibits a nonclassical magnetization plateau at one-third of the saturation magnetization $M_{\rm s}$~\cite{Nishimori,Chubokov,Nikuni,Honecker,Alicea,Farnell,Sakai,Hotta,Yamamoto1,Starykh2,Sellmann,Coletta}. With increasing magnetic field, successive QPTs from the Y-shaped coplanar state to the UUD state, and to the $2\,{:}\,1$ canted coplanar state occur~\cite{Chubokov,Nikuni,Honecker,Alicea,Farnell,Sakai,Hotta,Yamamoto1,Starykh2,Sellmann,Coletta}.

The quantum magnetization plateau characteristic of the 2D $S\,{=}\,1/2$ TLHAF was experimentally confirmed in Ba$_3$CoSb$_2$O$_9$~\cite{Shirata,Susuki,Prabhakaran}, in which magnetic Co$^{2+}$ ions with an effective spin $S\,{=}\,1/2$ at low temperatures~\cite{Abragam,Lines,Oguchi} form a uniform triangular lattice~\cite{Treiber_ZAAC1982,Doi}. However, the magnetization plateau was observed only for $H\,{\parallel}\,ab$ and not for $H\,{\parallel}\,c$~\cite{Susuki}. In addition, an unexpected high-field phase transition with a small magnetization jump was observed above the plateau state for $H\,{\parallel}\,ab$~\cite{Susuki}. This anisotropic magnetization process and the high-field phase transition were quantitatively explained by taking into account both the weak XY-like anisotropic exchange interaction and the weak antiferromagnetic interlayer exchange interaction~\cite{Gekht,Koutroulakis,Yamamoto2,Li}. 
The exchange anisotropy and antiferromagnetic interlayer exchange interaction also give rise to the different magnetic field--temperature phase diagrams for $H\,{\parallel}\,ab$ and $H\,{\parallel}\,c$~\cite{Koutroulakis,Zhou,Quirion}. 
Besides Ba$_3$CoSb$_2$O$_9$, 1/3-magnetization plateaux were observed in many quasi-2D quantum triangular-lattice antiferromagnets such as Cs$_2$CuBr$_4$~\cite{Ono1,Ono2,Fortune}, Ba$_2$CoTeO$_6$~\cite{Nick1}, Ba$_2$La$_2$CoTe$_2$O$_{12}$~\cite{Kojima}, CsYbSe$_2$~\cite{Xing}, NaYbSe$_2$~\cite{Ranjith,Ding} with the weak XY-like anisotropy, and Ba$_3$NiSb$_2$O$_9$~\cite{Shirata2}, Ba$_2$La$_2$NiTe$_2$O$_{12}$~\cite{Saito}, Na$_2$BaCo(PO$_4$)$_2$~\cite{Li3} with the weak Ising-like anisotropy. Observing a well-defined magnetization plateau shows that the sample has no exchange randomness, as it is theoretically shown that the exchange randomness makes indistinct the magnetization plateau~\cite{Watanabe}.

\begin{figure}[t]
\begin{center}
\includegraphics[width=0.95\linewidth]{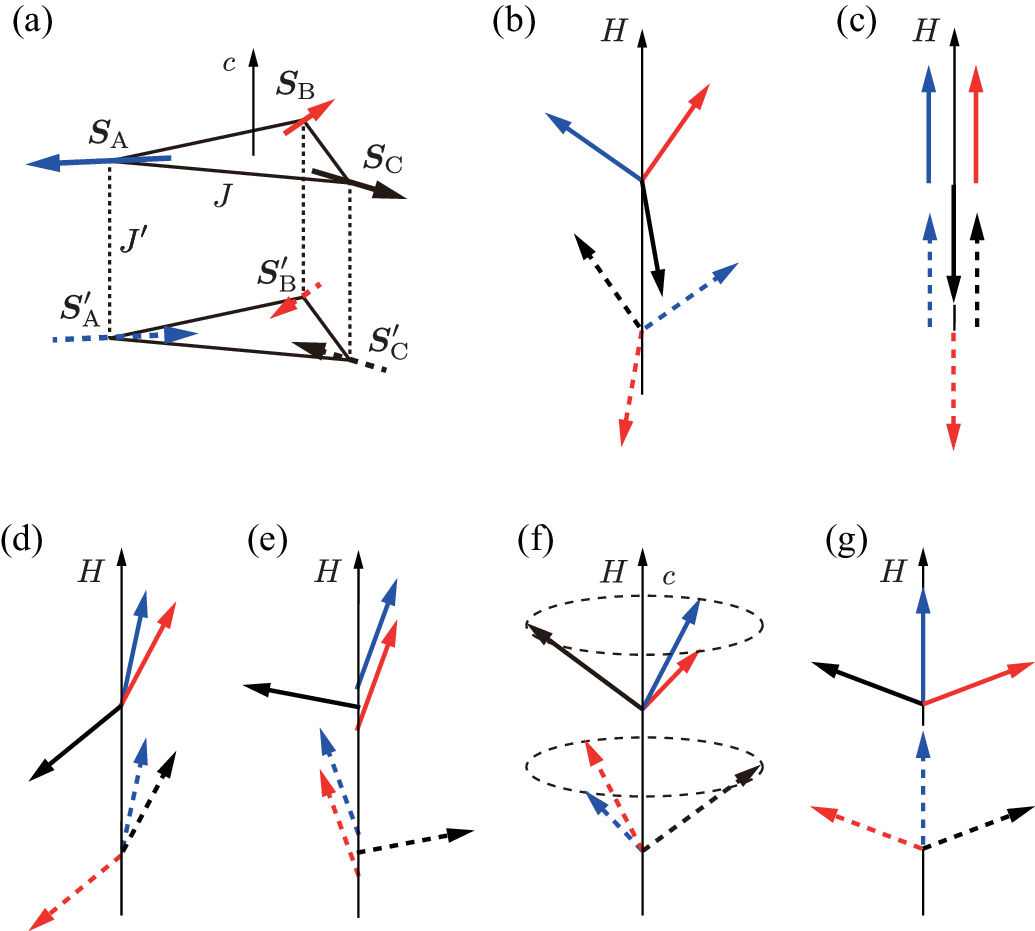}
\end{center}
\caption{(Color online) Possible spin structures of the $S\,{=}\,1/2$ triangular-lattice antiferromagnet with the weak XY-like anisotropy and the weak antiferromagnetic interlayer exchange interaction for $H\,{\parallel}\,ab$ and $H\,{\parallel}\,c$. Solid and dashed arrows respectively depict three sublattice spins (${\bm S}_{\rm A}, {\bm S}_{\rm B}, {\bm S}_{\rm C}$) and (${\bm S}_{\rm A}^{\prime}, {\bm S}_{\rm B}^{\prime}, {\bm S}_{\rm C}^{\prime}$) in neighboring triangular layers. (a) Zero-magnetic-field structure, (b) low-field coplanar structure for $H\,{\parallel}\,ab$, (c) up-up-down (UUD) structure, (d) and (e) lower and upper high-field coplanar structures, respectively, (f) umbrella structure for $H\,{\parallel}\,c$, and (g) $\Psi$-structure. Solid and dotted lines in (a) denote the intralayer and interlayer exchange interactions, respectively.
} 
\label{fig1}
\end{figure}

Figure~\ref{fig1} shows possible spin structures of the $S\,{=}\,1/2$ triangular-lattice antiferromagnet with the weak XY-like anisotropy and the weak antiferromagnetic interlayer exchange interaction at zero and finite magnetic fields for $H\,{\parallel}\,ab$ and $H\,{\parallel}\,c$. The QPT sequence for $H\,{\parallel}\,ab$ in Ba$_3$CoSb$_2$O$_9$ is (a)\,$\rightarrow$\,(b)\,$\rightarrow$\,(c)\,$\rightarrow$\,(d)\,$\rightarrow$\,(e)~\cite{Gekht,Koutroulakis,Yamamoto2,Li}, which was confirmed by a neutron diffraction experiment in high magnetic fields~\cite{Liu}. The transition from the lower high-field coplanar state (d) to the upper high-field coplanar state (e) is caused by the synergy between the weak antiferromagnetic interlayer exchange interaction and quantum fluctuation~\cite{Gekht,Koutroulakis,Yamamoto2,Li}. For $H\,{\parallel}\,c$, only a transition from the umbrella structure (f) to the lower-field coplanar structure (d) was observed in a previous study~\cite{Susuki}, although the theory predicts a similar high-field transition between the (d) and (e) structures as observed for $H\,{\parallel}\,ab$~\cite{Koutroulakis}. Later the high-field transition for $H\,{\parallel}\,c$ was observed via specific heat measurement~\cite{Fortune1}. The high-field transition from the (d) state to the (e) state is absent for a ferromagnetic interlayer exchange interaction as in the case of CsCuCl$_3$~\cite{Yamamoto2,Adachi,Tazuke,Nojiri_CsCuCl3,Tanaka_JPSJ1992,Motokawa,Stusser,Sera,Yamamoto4}, and the $2\,{:}\,1$ canted coplanar state is stable up to the saturation.

It is not clear how the field-induced QPTs change with varying field orientation in Ba$_3$CoSb$_2$O$_9$. Using the semiclassical theory, Koutroulakis {\it et al.}~\cite{Koutroulakis} investigated the field-orientation dependence of field-induced transitions and obtained a ground-state phase diagram for magnetic field vs. field orientation. The phase diagram appears nontrivial because the transition field related to the higher edge field of the 1/3-magnetization plateau for $H\,{\parallel}\,ab$ increases with the angle between the magnetic field and the $ab$ plane. However, the transition field connected to the lower edge field is almost independent of the field orientation, and the critical field for the high-field transition with a small magnetization jump has a clear maximum at an intermediate field angle. Recently, Fortune {\it et al.}~\cite{Fortune2} have reported the field-angle dependence of phase transitions in Ba$_3$CoSb$_2$O$_9$ measured via the field dependences of specific heat, thermal conductivity, and magnetic torque. However, the anomalies of these quantities were broad rather than sharp. Therefore, there appears to be a certain level of ambiguity in determining critical fields. It also seems difficult to identify whether these anomalies arise from the phase transition or crossover.

This paper presents the results of high-field magnetization measurements on Ba$_3$CoSb$_2$O$_9$ conducted by varying the angle $\theta$ between the magnetic field and the $c$ axis. We found a narrow 1/3-magnetization plateau and a high-field transition with a small magnetization jump for $H\,{\parallel}\,c$, which were not observed in the previous magnetization study~\cite{Susuki}. We compare obtained magnetic field--field angle ($H\,{-}\,{\theta}$) phase diagram with those obtained by the large-size cluster mean-field method combined with a scaling scheme  (CMF\,+\,S)~\cite{Yamamoto1,Yamamoto2} and the semiclassical theory~\cite{Koutroulakis}. The obtained $H\,{-}\,{\theta}$ phase diagram is generally in accordance with those obtained by these theories.

\section{Experimental details}
\begin{figure}[t]
\begin{center}
\includegraphics[width=0.6\linewidth]{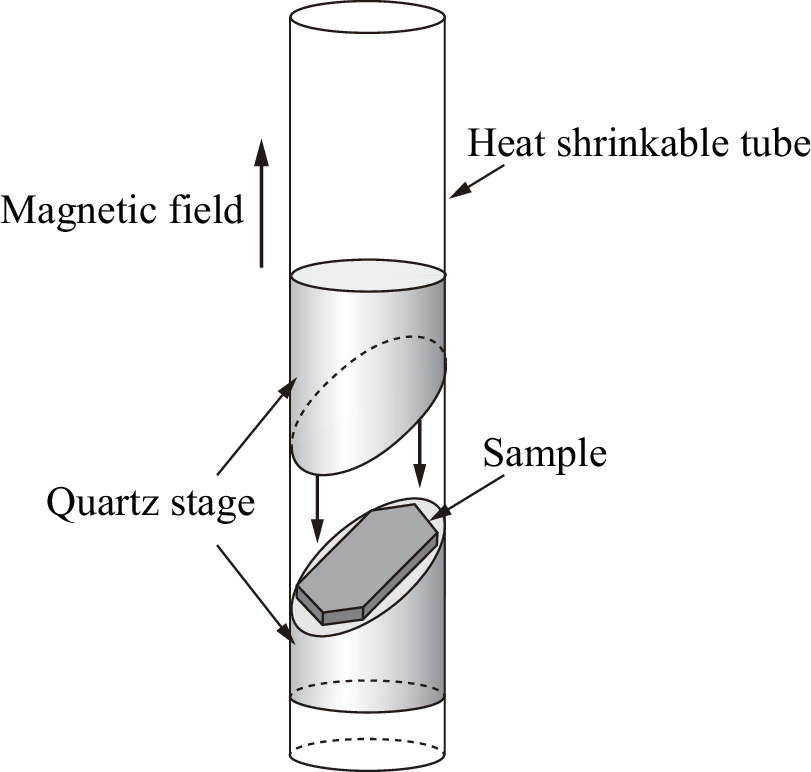}
\end{center}
\caption{(Color online) Sample holder used in the present high-field magnetization measurements. A plate-shaped sample is held between two quartz stages with a slope having an angle of ${\theta}$ to the horizontal plane.  
}
\label{fig2}
\end{figure}

Single crystals of Ba$_3$CoSb$_2$O$_9$ were grown from the melt using a Pt crucible or Pt tube with one end welded. Ba$_3$CoSb$_2$O$_9$ powder prepared by a procedure described in a previous paper~\cite{Shirata} was packed into the Pt crucible or Pt tube. The temperature of the furnace was decreased from 1700 to 1650\,$^{\circ}$C over two days. Plate-shaped crystals were extracted. The wide faces have already been identified to be the $ab$ plane by previous X-ray single-crystal diffraction and magnetization measurements.

The field-angle dependence of magnetization $M$ was measured at $T\,{=}\,1.8 $\,K, which is sufficiently lower than the N\'{e}el temperature $T_\mathrm{N}\,{=}\,3.8$\,K~\cite{Doi}, in magnetic fields up to $\mu_0H$\,{=}\,7\,T using a SQUID magnetometer (MPMS-XL, Quantum Design) equipped with a sample rotator. The magnetizations at $T\,{=}\,0.5 $\,K in magnetic fields up to $\mu_0H$\,{=}\,7\,T for $H\,{\parallel}\,ab$ and $H\,{\parallel}\,c$ were also measured using a $^3$He device (iHelium3, IQUANTUM).
The field-angle dependence of magnetization in pulsed high-field magnetic fields up to 52\,T was measured at $T$\,=\,1.3\,K at the Institute for Solid State Physics, University of Tokyo. Figure~\ref{fig2} illustrates the sample holder used in the high-field measurement. A plate-shaped sample with a wide $ab$ plane was held between two quartz stages with a slope having an angle of ${\theta}$ to the horizontal plane in a heat-shrinkable tube with an inside diameter of 2.5 mm. We prepared many sets of quartz stages with different slopes to measure the field-orientation dependence of the magnetization process. The magnetic field was applied parallel to the cylindrical axis of the holder, i.e., $H\,{\parallel}\,c$ when ${\theta}\,{=}\,0^{\circ}$. The absolute value of the high-field magnetization was calibrated with the magnetization measured by the SQUID magnetometer.

In the previous study~\cite{Susuki}, approximately ten single crystals were stacked in a holder. In the present experiment, we used one single crystal to prevent the dispersion of the crystal orientation. Consequently, sharp magnetization anomalies indicative of phase transitions were observed, as shown below. However, the absolute value of magnetization has a certain error level because the background signal, mainly due to the sample holder, is relatively enhanced.

\section{Results and discussion}


\begin{figure}[t]
\begin{center}
\includegraphics[width=0.95\linewidth]{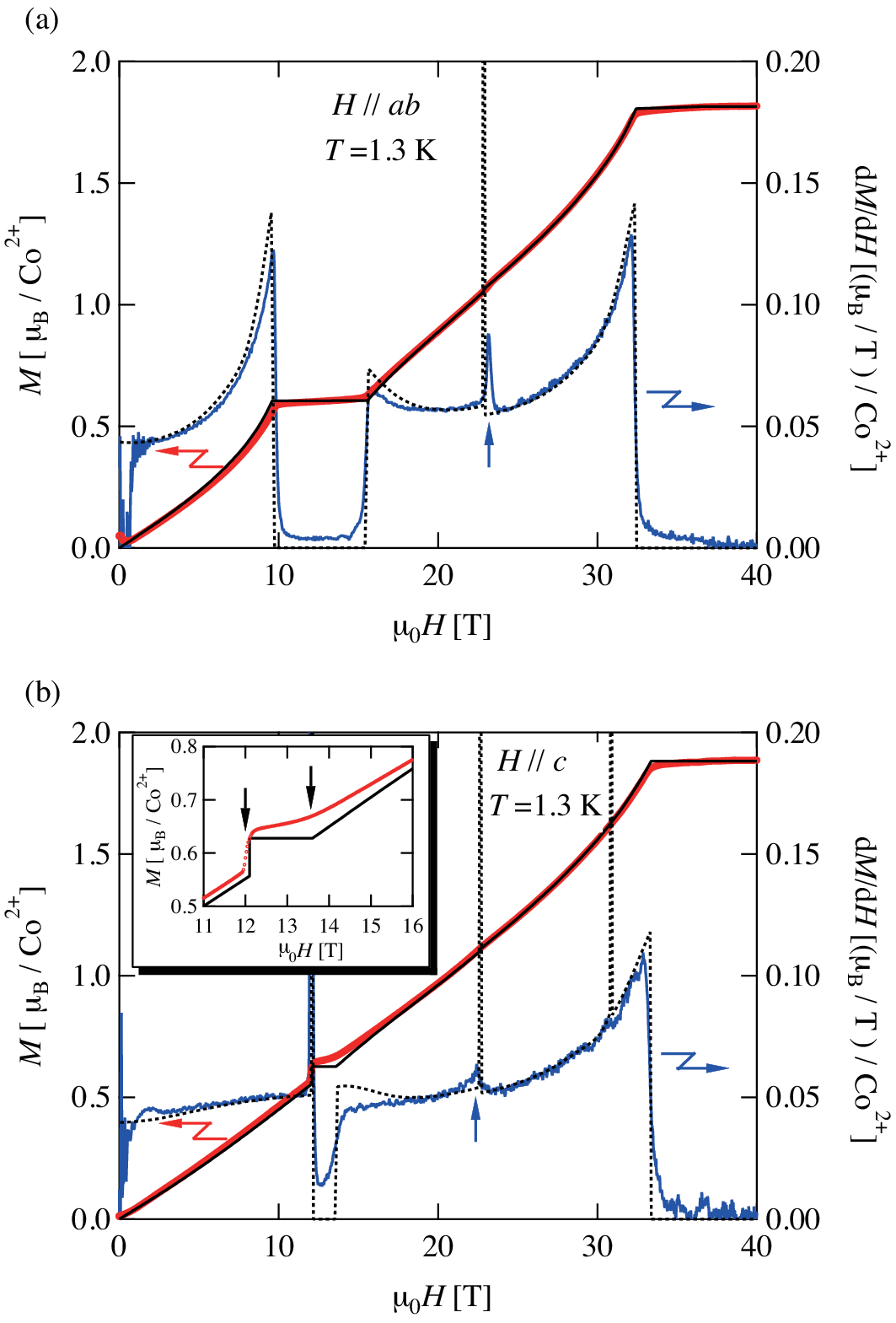}
\end{center}
\caption{(Color online) Field dependence of the magnetization $M$ (left) and its field derivative $dM/dH$ (right) of Ba$_3$CoSb$_2$O$_9$ measured at $T\,{=}\,1.3$\,K for (a) $H\,{\parallel}\,ab$ and for (b) $H\,{\parallel}\,c$, respectively, where the Van Vleck paramagnetism were corrected. Thin solid and dotted lines are theoretical magnetization curves and their field derivative, respectively, calculated by the CMF\,+\,S with $J_z/J\,{=}\,0.75$ and $J^{\prime}/J\,{=}\,0.05$, where the saturation fields and saturation magnetizations are normalized by the experimental values. The inset of (b) is an enlargement around the 1/3-magnetization plateau.  
}
\label{fig3}
\end{figure}

\begin{figure}[t]
\begin{center}
\includegraphics[width=0.95\linewidth]{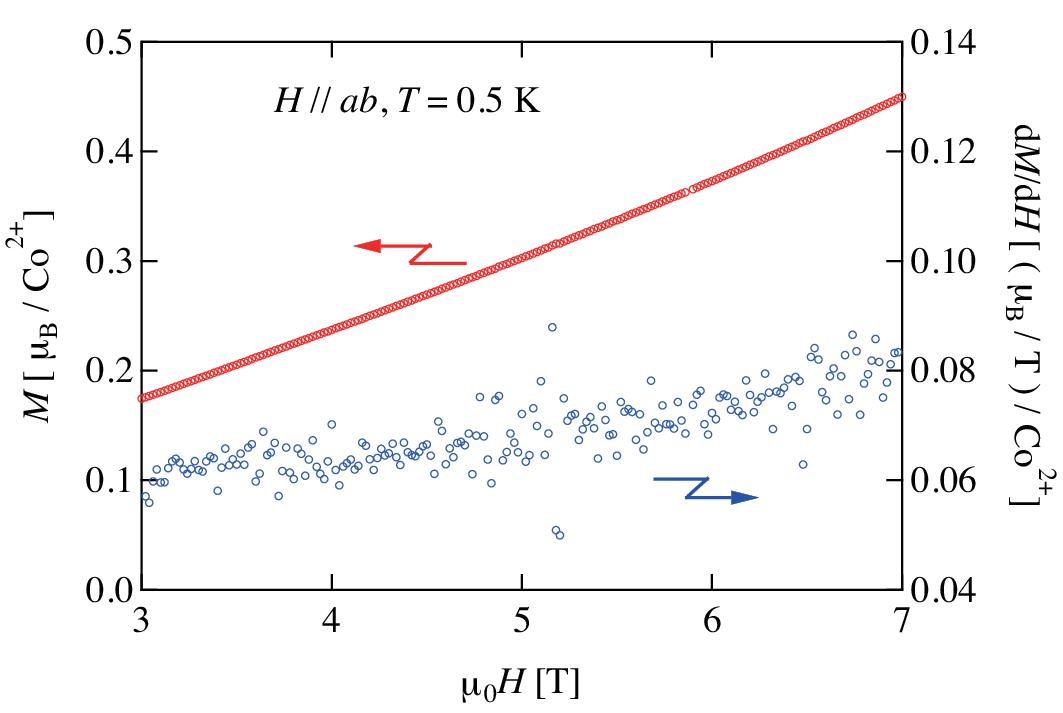}
\end{center}
\caption{(Color online) Low-field magnetization process of Ba$_3$CoSb$_2$O$_9$ measured at $T\,{=}\,0.5$\,K for $H\,{\parallel}\,ab$ using a SQUID magnetometer. 
}
\label{fig4}
\end{figure}

\begin{figure}[t]
\begin{center}
\includegraphics[width=0.95\linewidth]{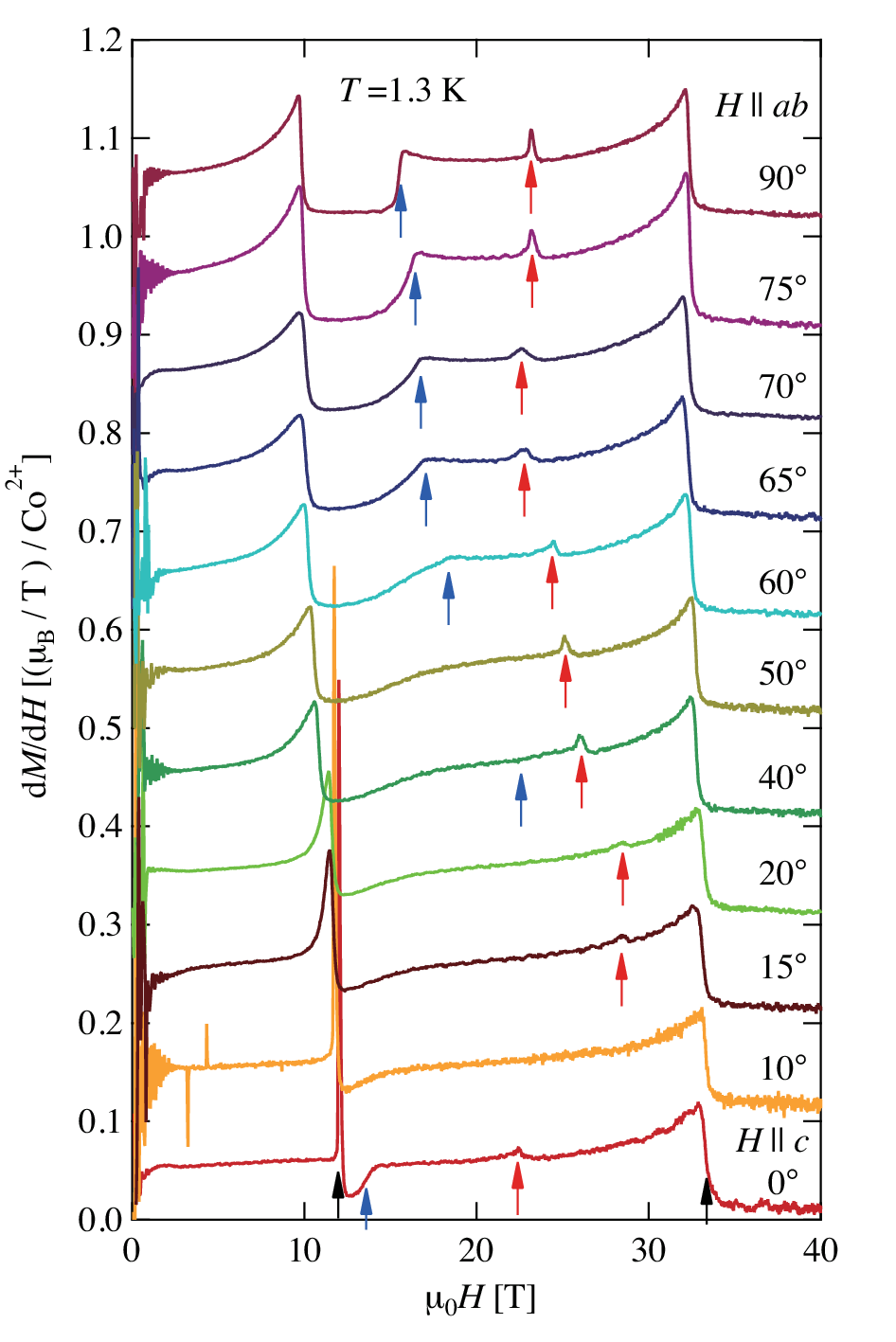}
\end{center}
\caption{(Color online) Field derivatives of raw magnetization ${\rm d}M/{\rm d}H$ vs magnetic field $H$ of Ba$_3$CoSb$_2$O$_9$ measured at various field angles ${\theta}$. The data are shifted upward by multiples of 0.1\,(${\mu}_{\rm B}$/T)/Co$^{2+}$. Vertical arrows denote transition fields. 
}
\label{fig5}
\end{figure}

Figures~\ref{fig3}\,(a) and (b) show the magnetization processes in Ba$_3$CoSb$_2$O$_9$ measured for $H\,{\parallel}\,ab$ and $H\,{\parallel}\,c$, respectively, at $T\,{=}\,1.3$\,K using pulsed magnetic fields~\cite{Comment}. The magnetization data were corrected for the Van Vleck paramagnetism, which was estimated from the magnetization slope above the saturation. The transition anomalies in $dM/dH$ for both field directions are much sharper than those observed in the previous study~\cite{Susuki}. This is because the dispersion of the crystal orientation is almost absent owing to the use of one single crystal. Field-induced QPTs and the saturation are identified at $H^{(ab)}_{\rm c1}{=}\,9.8$, $H^{(ab)}_{\rm c2}{=}\,15.6$, $H^{(ab)}_{\rm c3}{=}\,23.2$, and $H^{(ab)}_{\rm s}{=}\,32.4$\,T for $H\,{\parallel}\,ab$, and at $H^{(c)}_{\rm c1}{=}\,12.0$, $H^{(c)}_{\rm c2}{=}\,13.6$, $H^{(c)}_{\rm c3}{=}\,22.4$, and $H^{(c)}_{\rm s}{=}\,33.3$\,T for $H\,{\parallel}\,c$. It is particularly notable that a narrow 1/3-magnetization plateau and a transition with a small magnetization jump are clearly observed for $H\,{\parallel}\,c$ between $H^{(c)}_{\rm c1}$ and $H^{(c)}_{\rm c2}$, and at $H_{\rm c3}^{(c)}$, respectively, which were not observed in the previous magnetization study~\cite{Susuki}. This result indicates that the QPT sequence for $H\,{\parallel}\,c$ is (a)\,$\rightarrow$\,(f)\,$\rightarrow$\,(c)\,$\rightarrow$\,(d)\,$\rightarrow$\,(e) in the spin structures shown in Fig.~\ref{fig1}~\cite{Yamamoto1,Sellmann,Koutroulakis}. Note that this QPT sequence for $H\,{\parallel}\,c$ was also confirmed by a neutron diffraction experiment in high magnetic fields~\cite{Liu2}.

Here, we analyze the magnetization processes for $H\,{\parallel}\,ab$ and $H\,{\parallel}\,c$ based on a model expressed as
\begin{eqnarray}
{\cal H}&=&\sum_{\langle i,j\rangle}\left[J\left(S_i^xS_j^x + S_i^yS_j^y\right) + J_zS_i^zS_j^z\right] \nonumber\\ &+& J^{\prime}\sum_{\langle i,j\rangle^{\prime}} {\bm S}_i\cdot{\bm S}_j - g{\mu}_{\rm B}\sum_i {\bm H}\cdot{\bm S}_i,
\label{model}
\end{eqnarray}
where the first and second terms are the antiferromagnetic exchange interactions in a triangular layer and between triangular layers, respectively. We describe the dominant intralayer exchange interaction by the XXZ model. However, we express the weak interlayer exchange interaction by the Heisenberg model for simplification. We calculate the magnetization process in the ground state using the CMF\,{+}\,S method~\cite{Yamamoto1,Yamamoto2}. Thin solid and dotted lines in Figs.~\ref{fig3}(a) and (b) are theoretical magnetization and its field derivative, respectively, calculated with $J_z/J\,{=}\,0.75$ and $J^{\prime}/J\,{=}\,0.05$, where the saturation fields and saturation magnetizations for both field directions are normalized by the experimental values. For $H\,{\parallel}\,ab$, the agreement between experimental results and theory is almost perfect. However, for $H\,{\parallel}\,c$, there are some disagreements between experimental results and theory. Due to the finite temperature effect, the magnetization has a finite slope in the 1/3-magnetization plateau state, and the magnetization anomaly around the higher edge field of the plateau is not sharp compared to the theory. The theory predicts a transition at 30.9\,T between the upper high-field coplanar state and the highly symmetric $\Psi$-state shown in Figs.~\ref{fig1}\, (e) and (g), respectively. The $\Psi$-state is stabilized immediately below the saturation for $H\,{\parallel}\,c$ owing to the XY-like anisotropy and quantum fluctuation in a specific parameter range~\cite{Yamamoto1,Yamamoto2}. However, no apparent magnetization anomaly indicative of a transition between the upper high-field coplanar state and the $\Psi$-state was observed in the present experiment. Note that the emergence of the $\Psi$-state in the theoretical prediction for $H\,{\parallel}\,c$ depends largely on the value of the interlayer interaction $J^\prime$, as in the case of $H\,{\parallel}\,ab$~\cite{Yamamoto2}. For smaller $J^\prime/J$, say $0.025$, the $\Psi$-state below the saturation is absent, while the overall feature of the experimental magnetization curve is still well reproduced with the setting of a larger anisotropy, $J_z/J\,{=}\,0.73$. Further measurements are necessary to confirm the presence or absence of the transition.

Recently, Fortune {\it et al.}~\cite{Fortune2} have reported an additional low-magnetic-field phase transition at $H\,{\simeq}\,6$\,T for $H\,{\parallel}\,a$, which emerges only below $T\,{=}\,1$\,K before reaching the 1/3-magnetization plateau state. To confirm this transition, we measured the magnetization in Ba$_3$CoSb$_2$O$_9$ for $H\,{\parallel}\,ab$ at $T\,{=}\,0.5$\,K in a steady magnetic field up to 7\,T using a SQUID magnetometer. We did not identify the field direction in the $ab$ plane, because the in-plane magnetic anisotropy should be negligible. Figure~\ref{fig4} shows the magnetization $M$ (left) and its field derivative $dM/dH$ (right) as a function of $H$ at $T\,{=}\,0.5$\,K. No anomaly is observed around $H\,{=}\,6$\,T in both the $M$ and $dM/dH$ data. Thus, the origin of the thermodynamical anomalies at $H\,{\simeq}\,6$\,T for $H\,{\parallel}\,a$~\cite{Fortune2} is unclear at present.

\begin{figure}[t]
\begin{center}
\includegraphics[width=0.95\linewidth]{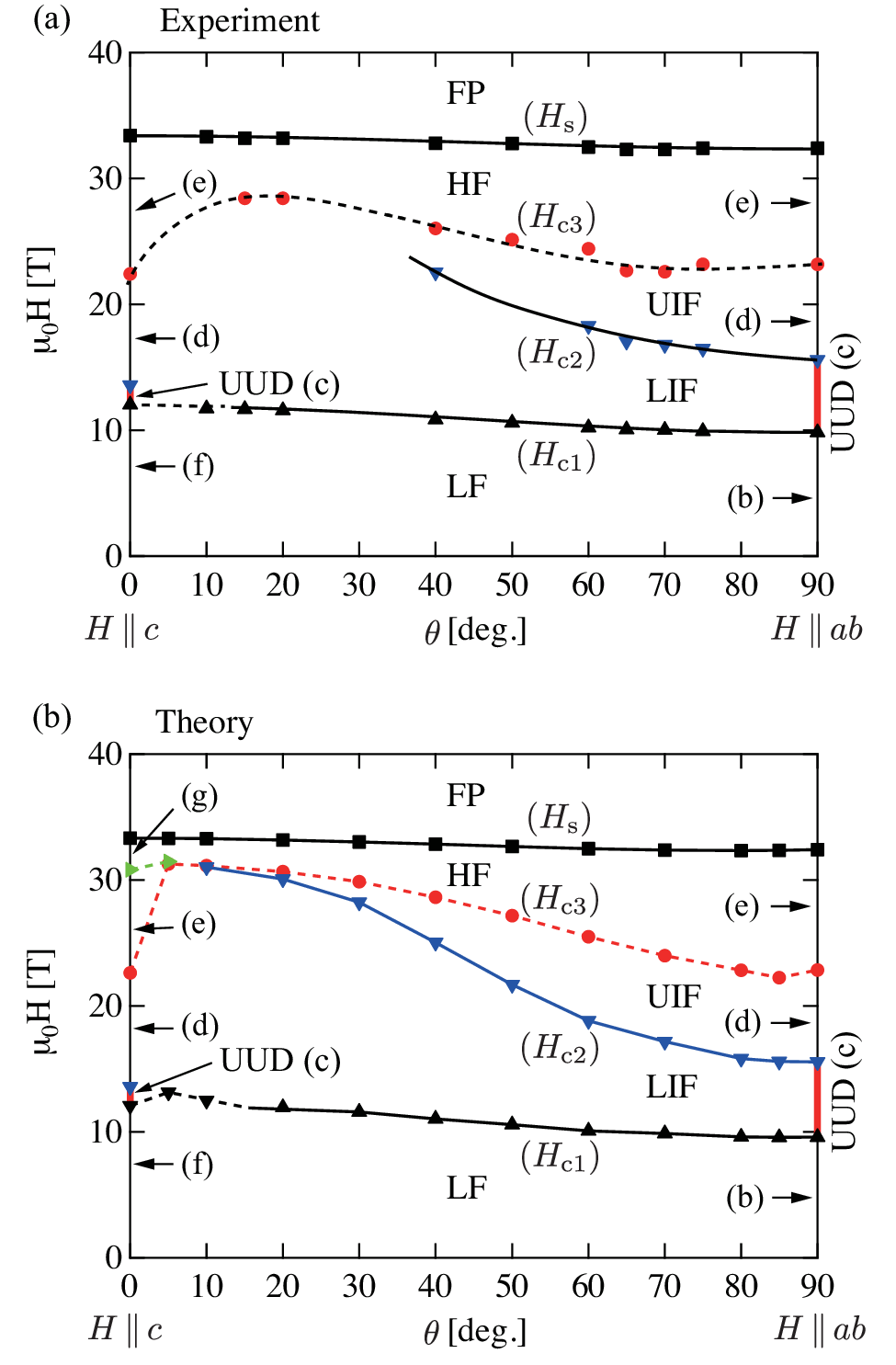}
\end{center}
\caption{(Color online) (a) Experimental and (b) theoretical ground-state phase diagrams parameterized by magnetic field and field angle ${\theta}$ ($H\,{-}\,{\theta}$ phase diagram) for Ba$_3$CoSb$_2$O$_9$ obtained by the magnetization measurements and the calculations using the CMF\,+\,S with $J_z/J\,{=}\,0.75$ and $J^{\prime}/J\,{=}\,0.05$, respectively. The saturation fields in (b) are normalized by the experimental saturation fields. Solid and dashed curves are guides to the eye, which denote the second-order and first-order phase transitions, respectively. Phases are labeled in accordance with Ref.~\cite{Koutroulakis} as the low-field phase (LF), lower-intermediate-field phase (LIF), upper-intermediate field (UIF), high-field phase (HF), and fully polarized phase (FP). $H_{\rm c1}$, $H_{\rm c2}$, $H_{\rm c3}$, and $H_{\rm s}$ are the critical fields for the LF--LIF, LIF--UIF, and UIF--HF (LIF--HF) transitions and saturation, respectively. Two 1/3-magnetization plateau (UUD) phases exist for $H\,{\parallel}\,ab$ and $H\,{\parallel}\,c$, depicted by two vertical red lines.} 
\label{fig6}
\end{figure}

\begin{figure}[t]
\begin{center}
\includegraphics[width=0.95\linewidth]{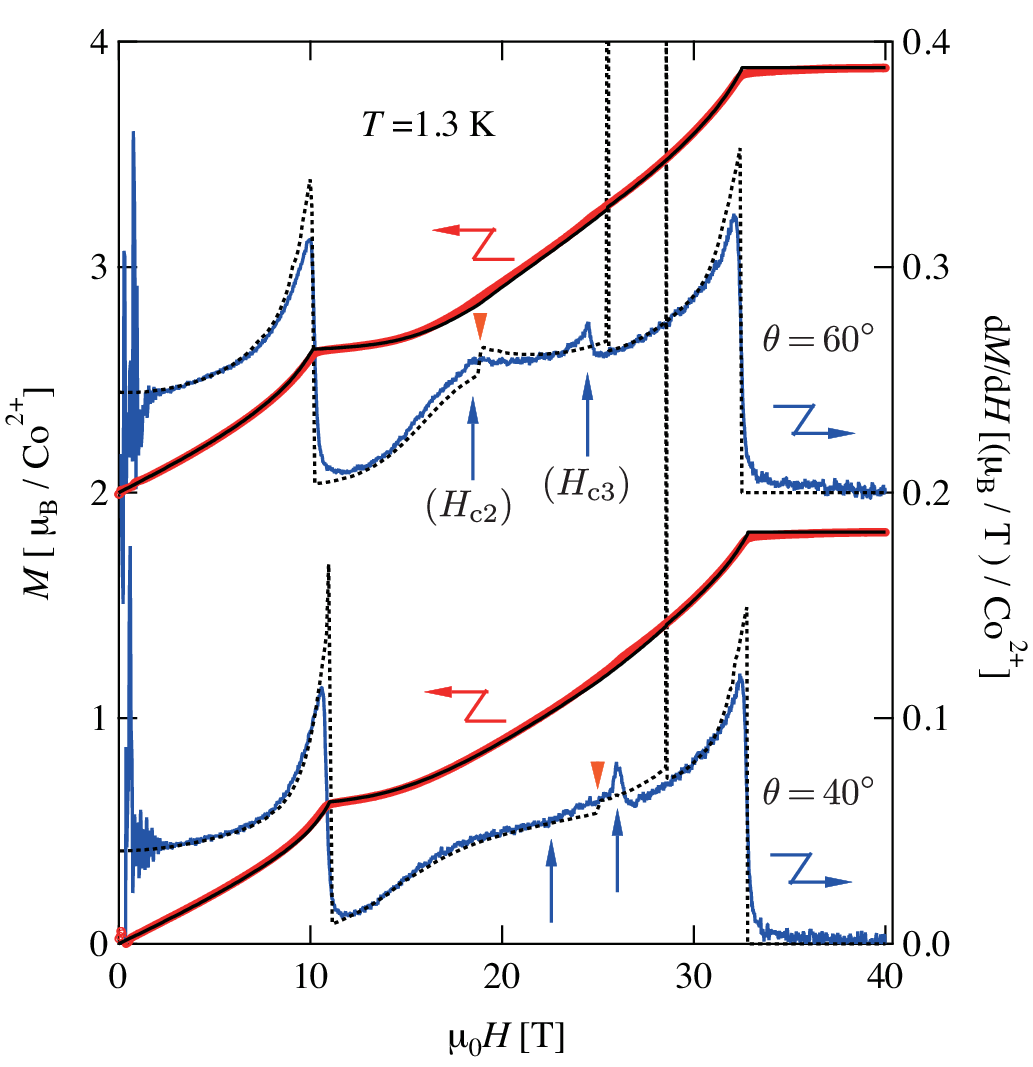}
\end{center}
\caption{(Color online) Field dependence of the magnetization $M$ (left) and its field derivative $dM/dH$ (right) of Ba$_3$CoSb$_2$O$_9$ measured at $T\,{=}\,1.3$\,K for ${\theta}\,{=}\,40^{\circ}$ and $60^{\circ}$. The data of $M$ and $dM/dH$ for ${\theta}\,{=}\,60^{\circ}$ are shifted by 2.0\,${\mu}_{\rm B}/{\rm Co}^{2+}$ and 0.2\,$({\mu}_{\rm B}/{\rm T})/{\rm Co}^{2+}$, respectively.
Thin dotted lines are $dM/dH$ vs $H$ calculated by the CMF\,+\,S with $J_z/J\,{=}\,0.75$ and $J^{\prime}/J\,{=}\,0.05$, where the saturation fields and saturation magnetizations are normalized by the experimental values. Vertical blue arrows and inverse orange triangles denote experimental and theoretical transition fields, respectively. 
}
\label{fig7}
\end{figure}

Figure~\ref{fig5} shows field derivatives of magnetization ${\rm d}M/{\rm d}H$ vs magnetic field $H$ measured at various field angles ${\theta}$. Field-induced QPTs are clearly observed as sharp anomalies of ${\rm d}M/{\rm d}H$. A systematic change in transition field can be seen upon varying the field angle ${\theta}$ from 90$^{\circ}$ to 0$^{\circ}$. Figure~\ref{fig6}\,(a) shows the transition data. We label the emerging phase in accordance with Ref.~\cite{Koutroulakis}. The low-field (LF) phase emerges at low magnetic fields and corresponds to the low-field coplanar state (b) and the umbrella state (f) below the 1/3-magnetization plateau for $H\,{\parallel}\,ab$ and $H\,{\parallel}\,c$, respectively. The lower-intermediate-field (LIF) and the upper-intermediate-field (UIF) phases continuously connect the 1/3-magnetization plateau state (UUD) and the lower-high-field state (d) for $H\,{\parallel}\,ab$, respectively. The high-field (HF) phase corresponds to the upper-high-field state (e) for $H\,{\parallel}\,ab$ and $H\,{\parallel}\,c$. In Fig.~\ref{fig6}\,(a), $H_{\rm c1}$, $H_{\rm c2}$, $H_{\rm c3}$, and $H_{\rm s}$ are the critical fields for the LF--LIF, LIF--UIF, and UIF--HF (LIF--HF) transitions and saturation, respectively.

The lowest critical field $H_{\rm c1}$, which is the lower edge field of the 1/3-magnetization plateau for $H\,{\parallel}\,ab$, slightly increases with decreasing ${\theta}$ from 90$^{\circ}$. For ${\theta}\,{\geq}\,15^{\circ}$, the anomaly of ${\rm d}M/{\rm d}H$ at $H_{\rm c1}$ is sharp but continuous, while for ${\theta}\,{=}\,10^{\circ}$ and 0$^{\circ}$, ${\rm d}M/{\rm d}H$ displays a $\delta$-function-like peak at $H_{\rm c1}$ indicative of the first-order transition with a magnetization jump. This result shows that the LF--LIF transition changes from the second-order to the first-order between 15$^{\circ}$ and 10$^{\circ}$. The change in the nature of the LF--LIF transition can be confirmed by the microscopic calculation using CMF\,+\,S method, as shown in Fig.~\ref{fig6}\,(b).

In the 1/3-magnetization plateau state for $H\,{\parallel}\,ab$, ${\rm d}M/{\rm d}H$ exhibits sharp rectangular anomalies at both edge fields $H_{\rm c1}$ and $H_{\rm c2}$. With decreasing field angle ${\theta}$ from 90$^{\circ}$, the anomaly of ${\rm d}M/{\rm d}H$ at $H_{\rm c1}$ remains sharp and vertical, irrespective of ${\theta}$, while the anomaly at $H_{\rm c2}$ is smoothed down to a cusp, as shown in Fig.~\ref{fig7}. This indicates that the 1/3-magnetization plateau at ${\theta}\,{=}\,90^{\circ}$ changes its shape to a ski-jump ramp with decreasing field angle ${\theta}$, which is consistent with the semiclassical theory~\cite{Koutroulakis}, and can also be reproduced by the microscopic calculation shown in Fig.~\ref{fig7}.

Note that the critical field $H_{\rm c2}$ for the LIF--UIF transition, which corresponds to the higher edge field of the 1/3-magnetization plateau for $H\,{\parallel}\,ab$, increases with decreasing ${\theta}$ from 90$^{\circ}$, thus increasing the field range between $H_{\rm c1}$ and $H_{\rm c2}$. 
Figure~\ref{fig7} shows the experimental and calculated field dependences of $dM/dH$ for ${\theta}\,{=}\,60^{\circ}$ and $40^{\circ}$. The experimental results are well reproduced by the CMF\,{+}\,S calculations with the same exchange parameters used for the calculations for $H\,{\parallel}\,ab$ and $H\,{\parallel}\,c$ shown in Figs.~\ref{fig3}\,(a) and (b). The transition anomaly at $H_{\rm c2}$ becomes small with decreasing ${\theta}$. 
The anomaly of experimental $dM/dH$ at $H_{\rm c2}$ is barely distinguishable at ${\theta}\,{=}\,40^{\circ}$ by comparing to the theoretical $dM/dH$, which exhibits a small stepwise anomaly at $H_{\rm c2}$. The calculated transition fields $H_{\rm c2}$ and $H_{\rm c3}$ are somewhat higher than those observed. As shown in Fig.~\ref{fig6}\,(b), the calculated LIF--UIF boundary meets the UIF--HF boundary. However, it is not clear whether the experimental LIF--UIF boundary has an endpoint or whether it meets the UIF--HF boundary. 

The two plateau states for $H\,{\parallel}\,ab$ and $H\,{\parallel}\,c$ are not the two ends of a plateau phase. These results are somewhat surprising because it is naively expected that a plateau phase exists for $0^{\circ}\,{<}\,{\theta}\,{<}\,90^{\circ}$ that connects the two plateau states for $H\,{\parallel}\,ab$ and $H\,{\parallel}\,c$, and that its field range decreases monotonically with decreasing ${\theta}$ from 90$^{\circ}$ to 0$^{\circ}$. It is considered that the enhanced field range of the LIF phase is attributed to the XY-like exchange anisotropy, which breaks the rotational symmetry around the external magnetic field~\cite{Koutroulakis,Yamamoto2}. This $U(1)$ symmetry is conserved only for $H\,{\parallel}\,c$.

The critical field $H_{\rm c3}$ for the UIF--HF (LIF--HF) transition shows a shallow minimum at ${\theta}\,{=}\,65{-}70^{\circ}$ and a clear maximum at ${\theta}\,{=}\,15{-}20^{\circ}$, as predicted by the semiclassical theory~\cite{Koutroulakis}. As shown in Fig.~\ref{fig3}\,(b), the CMF\,{+}\,S calculation for $H\,{\parallel}\,c$ with $J_z/J\,{=}\,0.75$ and $J^{\prime}/J\,{=}\,0.05$ predicts the emergence of the $\Psi$-state between the upper high-field coplanar state and the fully polarized state. The calculation shows that the LIF--UIF boundary connects with the transition field between the upper high-field coplanar state and the $\Psi$-state, as shown in  Fig.~\ref{fig6}\,(b). Further measurements are necessary to verify the theory. The $\Psi$-state is stabilized for $H\,{\parallel}\,c$ with the help of the XY-like anisotropy and quantum fluctuation~\cite{Yamamoto1,Yamamoto2}. Because the XY-like anisotropy in Ba$_2$La$_2$CoTe$_2$O$_{12}$ is relatively larger than that in Ba$_3$CoSb$_2$O$_9$~\cite{Kojima}, we can expect the emergence of the $\Psi$-state in Ba$_2$La$_2$CoTe$_2$O$_{12}$.

The high-field LIF--HF transition ($H_{\rm c3}$) is no longer observed in the ${\rm d}M/{\rm d}H$ data at ${\theta}\,{=}\,10^{\circ}$, as shown in Fig.~\ref{fig5}. At ${\theta}\,{=}\,10^{\circ}$, the anomaly assigned to $H_{\rm c2}$ is also absent. The narrow 1/3-magnetization plateau state at ${\theta}\,{=}\,0^{\circ}$ immediately changes into the shape of a ski-jump ramp at ${\theta}\,{=}\,10^{\circ}$. The CMF\,{+}\,S calculations show that the 1/3-magnetization plateau state emerges only for ${\theta}\,{<}\,5^{\circ}$, as shown in Fig.~\ref{fig6}\,(b), and that in the vicinity of ${\theta}\,{=}\,0^{\circ}$, the transition anomalies for $H_{\rm c2}$ and $H_{\rm c3}$ become so small that they are difficult to be detected. Thus, the phase boundaries near ${\theta}\,{=}\,0^{\circ}$ are sensitive to the field angle ${\theta}$. To investigate the phase boundaries near ${\theta}\,{=}\,0^{\circ}$ in detail, a precise magnetization measurement with the fine-tuning of the field angle ${\theta}$ is necessary.

In the previous magnetization study~\cite{Susuki}, neither the narrow 1/3-magnetization plateau nor the high-field transition was observed for $H\,{\parallel}\,c$. In that experiment, about ten single crystals were stacked in a cylindrical holder. Consequently, a dispersion of crystal orientations of at most ${\pm}10^{\circ}$ was inevitable. Thus, we deduce that the narrow 1/3-magnetization plateau and the high-field transition were smeared out owing to the dispersion of crystal orientations because the fields and the magnitude of anomalies at $H_{\rm c2}$ and $H_{\rm c3}$ are sensitive to the field angle ${\theta}$ near ${\theta}\,{=}\,0^{\circ}$.

\section{Conclusion}
The magnetic-field-orientation dependence of QPTs in the quasi-2D $S\,{=}\,1/2$ triangular-lattice antiferromagnet Ba$_3$CoSb$_2$O$_9$ has been investigated by magnetization measurements using one single crystal. The transition anomalies in the field derivative of the magnetization ${\rm d}M/{\rm d}H$ were sharp and clear because of the absence of the dispersion of the crystal orientation. We observed a 1/3-magnetization plateau and a high-field transition with a small magnetization jump for $H\,{\parallel}\,c$, which were not observed in the previous study~\cite{Susuki}. We obtained a ground-state $H\,{-}\,{\theta}$ phase diagram of Ba$_3$CoSb$_2$O$_9$ as shown in Fig.~\ref{fig6}\,(a). The features of the phase diagram are summarized as follows. The LF--LIF transition changes from the second-order to the first-order at ${\theta}\,{=}\,10{-}15^{\circ}$ with decreasing field angle ${\theta}$. The critical field $H_{\rm c2}$ for the LIF--UIF transition increases with decreasing ${\theta}$; thus, the field range of the LIF phase is enhanced. The critical field $H_{\rm c3}$ for the UIF(LIF)--HF transition has a clear maximum at ${\theta}\,{=}\,15{-}20^{\circ}$.
The $H\,{-}\,{\theta}$ phase diagram is generally in accordance with the theoretical phase diagram obtained by the CMF\,{+}\,S calculations with $J_z/J\,{=}\,0.75$ and $J^{\prime}/J\,{=}\,0.05$, as shown in Fig.~\ref{fig6}\,(b). Due to the small XY-like anisotropy and weak antiferromagnetic interlayer exchange interaction, the phase boundaries display nontrivial field-angle dependence. It will be of interest to investigate the $H\,{-}\,{\theta}$ phase diagram by tuning the anisotropy and interlayer exchange interaction using hydrostatic pressure as recently performed for CsCuCl$_3$~\cite{Sera,Yamamoto4}.

\section*{Acknowledgment}
We would like to thank O. Prokhnenko and X. Liu for showing us their experimental results of neutron diffraction in high magnetic fields and for the useful discussion. This work was supported by Grants-in-Aid for Scientific Research (A) (No.~17H01142) and (C) (No.~19K03711) from Japan Society for the Promotion of Science.


\begin{thebibliography}{25}
\bibitem{Comment} Although magnetization data for $H\,{\parallel}\,c$ has already been published in Ref.\,\cite{Liu2}, we show the data again because they are the basis for discussing the evolution of the critical fields with varying the field angle ${\theta}$.
\end{thebibliography}
\end{document}